\documentstyle[epsfig]{mn}
\newif\ifAMStwofonts
\AMStwofontstrue
\def\be{\begin{equation}}
\def\ee{\end{equation}}

\def\gtsima{$\; \buildrel > \over \sim \;$}
\def\ltsima{$\; \buildrel < \over \sim \;$}
\def\prosima{$\; \buildrel \propto \over \sim \;$}
\def\gsim{\lower.5ex\hbox{\gtsima}}
\def\lsim{\lower.5ex\hbox{\ltsima}}
\def\simgt{\lower.5ex\hbox{\gtsima}}
\def\simlt{\lower.5ex\hbox{\ltsima}}
\def\simpr{\lower.5ex\hbox{\prosima}}
\def\la{\lsim}
\def\ga{\gsim}

\title[Radio views of cosmic reionization]{Radio views of cosmic reionization }         
\author[Vald\'{e}s, Ciardi, Ferrara, Johnston-Hollitt \& R\"{o}ttgering]
{M. Vald\'{e}s$^1$, B. Ciardi$^2$, A. Ferrara$^1$, M. Johnston-Hollitt$^3$ \& H. R\"{o}ttgering$^4$\\
$^1$ SISSA/ISAS, via Beirut 2-4, 34014 Trieste, Italy. \\
$^2$ Max-Planck-Institut f\"ur Astrophysik, Karl-Schwarzschild-Stra\ss e 1, 85748 Garching, Germany.\\
$^3$ Discipline of Physics, University of Tasmania, Private Bag 21, Hobart, TAS 7005, Australia.\\
$^4$ Sterrewacht Leiden, Postbus 9513, 2300 RA Leiden, the Netherlands}

\date{December 2005}

\pagerange{\pageref{firstpage}--\pageref{lastpage}}
\pubyear{2005}
\begin{document}
\maketitle
\label{firstpage}
\begin{abstract}
We use numerical simulations of cosmic reionization and radiative processes related to the HI 21~cm emission line to produce 
synthetic radio maps as seen by next generation telescopes that will 
operate at low radio frequencies (e.g. LOFAR).
Two different scenarios, in which the end of reionization occurs 
early ($z\approx 13$) or late ($z\approx 8$) depending on the Initial Mass Function (IMF) of the 
first stars and ionizing photon escape fraction, have been explored. For each of these models we produce 
synthetic HI 21~cm emission maps by convolving the simulation outputs with the 
provisional LOFAR sampling function in the 
frequency range 76-140 MHz.  If reionization occurs late, LOFAR will be able to detect individual HI structures on arcmin 
scales, emitting at a brightness temperature of $\approx 35$~mK as a 3-$\sigma$ signal in about 
1000 hours of observing time. In the case of early reionization, the detection would be unlikely, due to 
decreased sensitivity and increased sky temperatures. 
These results assume that 
ionospheric, interference and foreground issues are fully under control.
\end{abstract}


\section{Introduction}
Cosmic recombination left the gas in the universe in a nearly uniform, cold 
and neutral state. This vast sea of baryonic material was only rippled by 
tiny density fluctuations which, under the action of gravity, 
later grew into bound objects inside which eventually stars formed. This 
evolutionary trend was however counteracted by a number of physical processes:
the finite, albeit low, Jeans mass scale ($M\approx 10^4 M_\odot$) preventing the
gas to closely track the collapse of the underlying dark matter component; the lack
of the standard cooling agents (heavy elements, dust) for the gas; the fragility 
and low cooling efficiency of available molecules (H$_2$, HD, LiH); finally, a number
of radiative and mechanical feedback effects (e.g. Ciardi \& Ferrara 2005) 
arising from the formation of the first luminous sources. As a result of these 
complications, shaping the universe as we currently monitor it up to the most
remote objects was a somewhat grueling process. Therefore, a non-negligible fraction 
($\approx$ 1\%) of cosmic time was spent in a relatively dark and quiescent state that,
following M. Rees suggestion, we currently denote as the Dark Ages of the universe.   

The investigation of the Dark Ages is still in its infancy, and very 
little, if any, experimental support to theory is currently available.
However, the most promising (by far) technique to explore the evolution of 
the intergalactic medium (IGM) during these remote times is the    
study of the 21~cm hyperfine triplet-singlet level transition of the ground state  
of neutral hydrogen. This line could, in principle, allow a superb 
tracing of the HI distribution in the early universe, and therefore a
reconstruction of the reionization history as governed by the first luminous sources. 

In recent years  
two independent sets of data coming from QSO absorption line experiments,  
and the analysis of the Cosmic Microwave Background (CMB) 
temperature-polarization cross-correlation spectrum, constrained the epoch of reionization 
(EoR).
In brief, the sharp raise of the Gunn-Peterson Ly$\alpha$ (and Ly$\beta$) optical
depth derived from the analysis of $z \gsim 5$ quasar spectra, has led some authors to 
interpret this result as an indication of the rapid neutral/ionized transition 
to take place at relatively low redshift
(e.g. White et al. 2003).
The first year \textit{Wilkinson Microwave Anisotropy Probe} (\textit{WMAP}) data instead seemed to indicate that either the
EoR was located at a higher redshift or that a more complex evolution
of the gas ionized fraction took place (e.g. Kogut et al. 2003). This interpretation has to be revised
in view of the newly released data (Spergel et al. 2006), which favor a lower reionization redshift.
In recent years a large effort has been made to numerically 
simulate the formation of the first structures and the resulting photoionization of the IGM
(e.g. Gnedin 2000; Ciardi, Ferrara, \& White 2003 (CFW); Sokasian et al. 2003; Gnedin 2004; Iliev et al. 2006).
Although it is possible to devise models which are in agreement with both sets of data (e.g. Choudhury 
\& Ferrara 2005; Gallerani, Choudhury \& Ferrara 2005), independent measurements are eagerly required.  

Redshifted 21~cm line emission from the Dark Ages represents a unique chance to 
fully map the spatial distribution of intergalactic hydrogen (e.g. 
Madau, Meiksin \& Rees 1997; Ciardi \& Madau 2003 (CM);
Furlanetto, Sokasian \& Hernquist 2004; Furlanetto, Zaldarriaga \& Hernquist 2004) 
as a function of redshift. 
The size of the structures that could be detected depends on the design of future radio telescopes. The next
generation of radio telescopes such as the Square Kilometer Array (SKA), the
LOw Frequency ARray (LOFAR), the PrimevAl Structure Telescope (PAST)
and the Mileura Wide-field Array (MWA) will probably have the sensitivity for HI
mapping at resolutions of the order of a few arcminutes (e.g. Pen, Wu, \& Peterson 2004; Bowman, Morales, \& Hewitt 2005;
Kassim et al. 2004; Wyithe, Loeb, \& Barnes 2005).
Our aim here is limited, in the sense that we will describe the likelihood of one particular instrument 
(i.e. LOFAR) to be able to detect a signal from EoR. To do this we have simulated
observations of the EoR signal for two very general (early, late) reionization scenarios
with an idealized LOFAR array, using state-of-art cosmological radiative transfer numerical computations. 
Although this is only a first step towards a proper modelling, the results already give 
a clear flavor of the tremendous potential that these types of experiments will provide
in the very near future.


\section{21~cm radio emission}
\label{21cmline}

The 21~cm line is associated with the hyperfine transition between the triplet and the
singlet levels of the hydrogen ground state.
This transition is governed by the spin temperature, $T_{S}$, defined as:
\begin{equation}
\frac{n_{1}}{n_{0}}=3 \exp \left(-\frac{{T_{\star}}}{T_{S}}\right),
\end{equation}
where $n_{0}$ and $n_{1}$ are the number densities of hydrogen atoms in the
singlet and triplet ground hyperfine levels, and $T_{\star}=0.068$~K is the 
temperature corresponding to the transition energy.

In the presence of the CMB radiation alone the spin temperature reaches thermal 
equilibrium with $T_{\rm CMB}=2.73(1+z)$~K on a short timescale, making the HI undetectable
in emission or absorption. A mechanism is then required to decouple $T_{S}$ from $T_{\rm CMB}$
to observe the line. For the range of densities typical of the IGM, the scattering by Ly$\alpha$ photons 
is an efficient mechanism, as it couples $T_{S}$ to the kinetic gas temperature $T_{K}$
mixing the hyperfine levels of the neutral hydrogen in its ground state through intermediate
transitions to the excited $2p$ state. This is the so called Wouthuysen-Field process, or 
Ly$\alpha$ pumping (e.g. Wouthuysen 1952; 
Field 1959; Hirata 2005). 
For the Ly$\alpha$ pumping to be effective the Ly$\alpha$ background intensity, $J_{\alpha}$, 
has to satisfy the condition (e.g. CM): 
\begin{equation}
J_{\alpha}\geq 9 \times 10^{-23}(1+z)\,\mbox{ergs}\,\,\mbox{cm}^{-2}\,\mbox{s}^{-1}\,
\mbox{Hz}^{-1}\, \mbox{sr}^{-1}, 
\end{equation}
at the redshift of interest.  CM find that this condition on the diffuse 
Ly$\alpha$ intensity 
is satisfied 
between $z\approx 20$ and the EoR.

The 21~cm radiation intensity can be expressed by the differential brightness temperature
between a neutral hydrogen patch and the CMB:
\begin{equation}
\delta T_{b}\,\simeq\,\frac{T_{S}-T_{CMB}}{1+z}\,\tau,
\end{equation}
where $\tau$ is the optical depth of the neutral IGM at $21(1+z)\,\,\mbox{cm}$:
\begin{equation}
\tau\,=\,\frac{3c^{3} h_{P} A_{10}}{32 \pi k {\nu_{0}}^{2} T_{S}H(z)}n_{\rm HI}.
\end{equation}
In eq. 4, $h_{P}$ and $k$ are the Planck and Boltzmann constants, respectively; $\nu_{0} = 1420$~MHz
is the 21~cm hyperfine transition frequency, $A_{10}=2.85 \times 10^{-15} {\mbox{s}}^{-1}$ is the
21~cm transition spontaneous decay rate, and $n_{\rm HI}$ is the local HI number density.

If $T_{K}$ (and thus $T_{S}$) is higher than $T_{\rm CMB}$ the neutral IGM will be visible in emission 
against the CMB; on the contrary, if $T_{S} < T_{\rm CMB}$ it will be visible in absorption.
The heating by primordial sources of radiation should easily push $T_{S}$ to values
$\gg T_{\rm CMB}$ at the redshifts of interest (see e.g. Madau, Meiksin \& Rees 1997;
Carilli et al. 2002; Venkatesan, Giroux \& Shull 2001; Chen \& Miralda-Escud\`e 2004).
Then the IGM will be observable in emission at a level that is independent of the exact
value of $T_{S}$ and is directly proportional to $n_{\rm HI}$.
A radio interferometer able to make a 21~cm tomography would allow to probe accurately the 
reionization history.


\section{Simulations of 21~cm line emission}

In this Section we briefly describe the simulations of 21~cm line emission devised to 
produce synthetic brightness temperature maps; we refer the reader to the original papers 
(Ciardi, Stoehr \& White 2003; CFW; CM) for more details.
The simulations of reionization\footnote{The simulations assume a $\Lambda$CDM ``concordance'' cosmology
with $\Omega_m$=0.3, $\Omega_{\Lambda}$=0.7, $h=$0.7, $\Omega_b$=0.04,
$n$=1 and $\sigma_8$=0.9. These parameters are within the \textit{WMAP}
experimental error bars (Spergel et al. 2003).}
employ a combination of high resolution
N-body simulations (to describe the distribution of dark
matter and diffuse gas; Springel, Yoshida, \& White 2001;
Yoshida, Sheth, \& Diaferio 2001), a semi-analytic model
of galaxy formation (to track gas cooling, star formation,
and feedback from supernovae; Kauffmann et al. 1999;
Springel et al. 2001) and the Monte Carlo radiative transfer code CRASH 
(to follow the propagation of ionizing photons into the IGM; Ciardi et al. 
2001; Maselli, Ferrara, \& Ciardi 2003). 
The linear box size is 20$h^{-1}$~Mpc comoving, which corresponds to
$10.59$~arcmin at $z=9.26$.

Simulations were run with two sets of parameters for the stellar emission properties.
The run labeled S5 ({\it late reionization} model) in CFW adopts
a population of PopIII stars with a Salpeter Initial Mass Function (IMF) and an
escape fraction of ionizing photons $f_{esc}=5$\%. The L20 ({\it early reionization}) run adopts
a mildly top-heavy Larson IMF and  $f_{esc}=20$\%. Complete reionization (i.e. an ionized fraction per unit volume $>$ 0.99) is
reached by $z \approx 8$ and $z \approx 13$, respectively.
The output of the simulation provides, among other quantities, a series of 128$^3$ arrays with gas number 
density and ionization fraction. These were used by CM to derive (from the equations described in
the previous section) the expected 21~cm emission from the IGM at different redshifts.


\section{Basic radioastronomy equations}

In this Section we will briefly introduce the concepts of radio astronomy relevant for this
study (for more radio astronomy details see Taylor, Carilli, Perley 1998).
The quantity measured by a radio telescope is the sampled visibility function:
\begin{equation}
V_{\nu}(u,\,v)S(u,\,v) = \int\!\!\!\!\int I^{D}_{\nu} (l,\,m) \exp[-2\pi i (ul+\nu m)]dl\,dm,
\end{equation}
where $(u,\,v)$ are the coordinates  
in the plane perpendicular to the viewing direction (Fourier or $(u,\,v)$ plane),
$\nu$ is the frequency of the observed 
monochromatic component, $l$ ($m$) is  the
director cosine measured with respect to $u$ ($v$) defining the object position in the sky,
and $S(u,\,v)$ is the sampling function (also called $(u,\,v)$ coverage), defined by the 
projection of the baselines vectors between the telescopes,
which accounts for the
fact that it is not possible to sample the visibility function on the whole $(u,\,v)$ plane.
The Fourier Transform of the sampled visibility function gives the dirty image:
\begin{equation}
{I^{D}_{\nu}} (l,\,m)= \int\!\!\!\!\int {V}_{\nu}(u,\,v)\,S(u,\,v) \exp[2\pi i (ul+\nu m)]du\,dv.
\end{equation}
The dirty image is equivalent to the actual distribution of the radiation intensity, $I_{\nu}$,
convolved with the dirty beam or point spread function (PSF). $V_{\nu}$ is measured 
in units of flux density (Jansky, Jy), while $I_{\nu}$ is a flux density per unit of solid angle
(e.g. Jy/beam area) that is often described in terms of brightness temperature (K)
through the conversion $T_b=c^2I_{\nu}/2k\,\nu$.

In a radio interferometer the monochromatic $(u,\,v)$ coverage is given by the projections on the ($u,\,v$) plane
of the baselines between the observing stations, such that 
\begin{equation}
\left(
\begin{array}{c}
u \\
v 
\end{array}
\right)=\frac{1}{\lambda} \,\,A
\left(
\begin{array}{c}
L_{x} \\
L_{y} 
\end{array}
\right),
\end{equation}
\begin{equation}
A=\left(
\begin{array}{cc}
\sin HA_{0} & \cos HA_{0} \\
-\sin \delta_{0} \cos HA_{0} & \sin \delta_{0} \sin HA_{0} \\
\end{array}
\right),
\end{equation}
where $HA_{0}$ and $\delta_{0}$ are the hour angle and declination of the phase center,
$\lambda$ is the mean band wavelength and $L$ is the length of the baselines.
Note that this approximation ignores any non-planar effects appropriate 
for small field imaging.


\section{LOFAR view of reionization}

LOFAR will explore the Universe at frequencies between 30 and 80 MHz (Low Band)
and between 110 and 240 MHz (High Band); 
it will be composed of 
thousands of antennas divided initially in 77 stations spread over a $\sim 100$ km radius,
and eventually in up to 100 stations, with a maximum baseline of 
$\sim 400$ km.
This could be later extended to more than 1000~km if, in addition to the Netherlands where LOFAR
is presently under construction, other European countries join the project.\footnote{See
e.g. www.mpifr-bonn.mpg.de/staff/wsherwood/LOFAR/
white.paper.oct6.pdf}  
The $\sim 400$ km baselines will guarantee a spatial 
resolution as high as $\approx 1$~arcsec at 150 MHz.
Several problems associated with low radio frequency observations need to be 
addressed, such as the ionospheric scintillation, interferences from human-generated 
radio signals and extended/point source foregrounds.
The sensitivity will be optimized for the reionization experiment by concentrating about 40\%
of the collecting area in the inner ($\approx 2$~km diameter) radius or \textit{Virtual Core} of the instrument. 
The LOFAR correlator is such that it may be configured in several different
modes to achieve differing numbers of beams on the sky and spectral resolutions. 
More specifically, the product between the number of channels, $N_{ch}$, 
and the number 
of polarizations is constant and equal to the total number of digital 
signal paths.
In addition, the total available bandwidth of 32~MHz can be divided typically into up to 8 different pieces
to give different beams on the sky. Thus trade offs can
be made between these parameters to optimize the array for the desired observations, i.e. bandwidth
may be traded for an increased number of beams or polarization can be traded for an increased number of
spectral channels. 

From the LOFAR provisory station configuration (100 stations with baselines of
up to $\sim 360$ km, J. Noordam, private communication) 
we have computed the $(u,\,v)$ coverage, which is extremely complete 
as can be seen from Figs. 1-2. Fig. 1, in particular, shows the instantaneous $(u,\,v)$
coverage for a declination $\delta_0=\pi /2$ while the box in the
lower right corner shows a zoom of the $(u,\,v)$ coverage between -5 and 5 Kilo-$\lambda$;
Fig. 2 reports the $(u,\,v)$ coverage for a 8-hour continuous observing
time at $\delta_0=\pi /6$. In this case the coverage is 
very dense thanks to the earth rotation synthesis and to the configuration of LOFAR's stations. 
\begin{figure}
\begin{center}
\includegraphics[width=7.8cm]{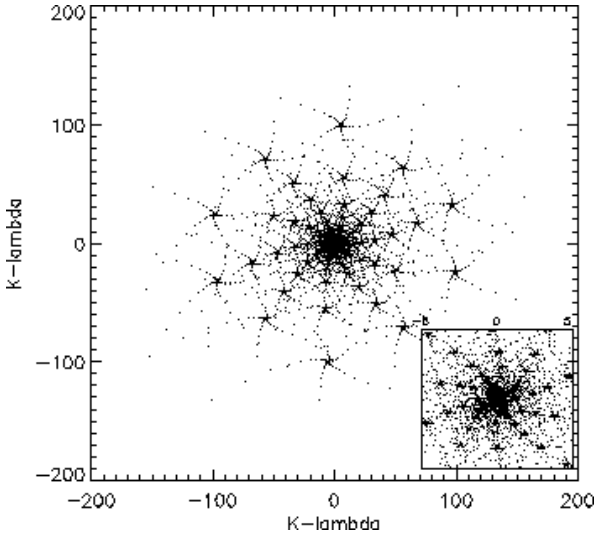}
\end{center}
\caption{LOFAR instantaneous $(u,v)$ coverage for a declination $\delta_0=\pi /2$. The units on the $x$ and $y$ axis
are Kilo-$\lambda$, where $\lambda$ is the mean band wavelength. The box on the lower right corner shows a
zoom of the $(u,v)$ coverage between -5 and 5 Kilo-$\lambda$.}
\label{graph1}
\end{figure}

\begin{figure}
\begin{center}
\includegraphics[width=7.8cm]{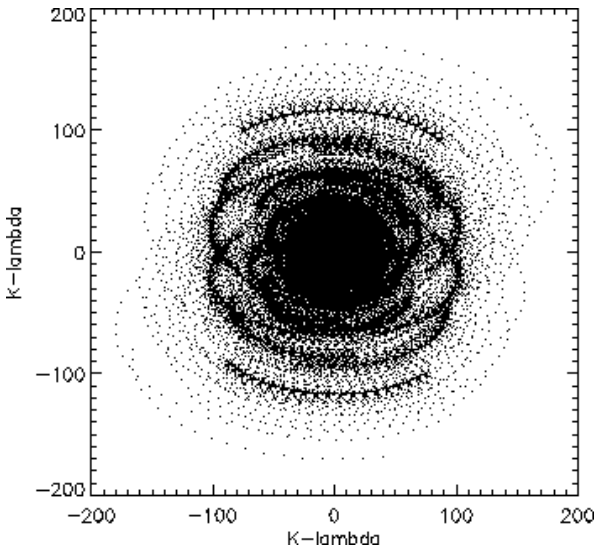}
\end{center}
\caption{LOFAR $(u,v)$ coverage for $\delta_0=\pi/6$ declination angle after 8 hours integration
time, with the same units as in Fig.1.}
\label{graph2}
\end{figure}


\subsection{Synthetic radio maps}

To produce synthetic radio maps of the reionization process as seen by LOFAR we proceed as
follows.
We denote by $l_c=20/128 \; h^{-1}$~Mpc the comoving thickness of a slice of the simulation box, and with:
\begin{equation}
L_c\approx 1.7 \left(\frac{\Delta \nu_{obs}}{0.1 \mbox{MHz}}\right) \left(\frac{1+z}{10}\right)^{1/2} \left(\frac{\Omega_m h^2}
{0.15}\right)^{-1/2} \; {\rm Mpc},
\end{equation}
the comoving length corresponding to an observational bandwidth $\Delta \nu_{obs}$.
The number of slices that corresponds to a
fixed $\Delta \nu_{obs}$ is $n_c=L_c/l_c$. 
For a given $\Delta \nu_{obs}$, we extract from each simulation a 2D map of brightness temperature 
by averaging the values of $\delta T_b$ on a number of slices $n_c$. 

The maps are first Fourier-transformed and then 
they are convolved with the $(u,\,v)$ coverage, $S(u,\,v)$, 
given in Eqs. 7-8 and shown in Fig. 2. 
Such convolution is necessary to take into account 
the limitations due to LOFAR's sampling function. 
We assume a monochromatic $(u,\,v)$ coverage, i.e. we neglect the width of
the tracks in the $(u,\,v)$ plane. This represents a simplification,
however we find that for the present study this approximation has a small impact on the results in 
comparison to the sensitivity limits of the instrument, because of the 
exceptionally complete $(u,\,v)$ coverage due to the log-spiral arms configuration of LOFAR.
Next, an inverse Fourier transform has been performed to return to the image 
plane including the estimated instrumental sensitivity on different
scales and frequencies as given at the official LOFAR website \footnote{http://www.lofar.org/science}. The 
sensitivity of an interferometer is $\propto (\Delta \nu_{obs} \Delta t)^{-1/2}$; we assumed 
an observational bandwidth $\Delta\nu_{obs} =128$~kHz and an observation time $\Delta t =1000$~hours.\footnote{We note 
that a total observing time of 1000 hours 
can only be obtained through a series of observations taking place 
during a few months. There will be slight shifts in $(u,\,v)$ coverages 
between the different observations, resulting in an ever better $(u,\,v)$ coverage 
as sketched for a single 8 hour observations.}
Finally, we have convolved the resulting image with a gaussian beam of $\approx 3$ arcmin. This choice represents the
best compromise between LOFAR sensitivity (which decreases for smaller beams) and the size of the simulation 
box, which is limited to $\approx 11$~arcmin. On this angular scale LOFAR will likely be able to detect 
the 21~cm signal thanks to its large collecting area in the Virtual Core. 
We note that this is mathematically analogous to applying a taper in the $(u,\,v)$ plane
 to give greater weighting to data from the shorter baselines, which is entirely consistent with the most
critical part of the data being collected by the Virtual Core. 

The results of this procedure are presented in Fig. 3 for the S5 model, i.e. the {\it late reionization}  
scenario in which the EoR is located at $z\approx 8$, the most promising in terms of 21~cm line detection 
from the pre-reionization IGM.  The maps represent the 21~cm emission brightness temperature in logarithmic scale,
both directly from the numerical simulation (i.e. before the convolution with LOFAR instrumental
characteristics) and in the synthetic LOFAR images;
the calculations were performed using the simulation outputs at redshifts 
$z=$ 10.6, 9.89 and 9.26 (corresponding to $\nu=$ 122, 130 and 138 MHz respectively).
We find a region that should be detectable at a 3-$\sigma$ level, 
showing that if reionization occurred relatively late LOFAR should detect the signal from the neutral IGM.
The resolution and sensitivity of the instrument allow the identification not only of the largely neutral 
IGM, but also of the HII holes produced by the ionizing sources. 
The brightest emission peaks at $z=10.6$ have $\delta T_b = 35$~mK (the corresponding value
in the original simulated map is $\sim 0.2$~K),
whereas the holes are typically a factor of
5 below that - a dynamical range large enough to allow their robust
identification as LOFAR's sensitivity will be $\sim 10$~mK at these scales,
frequencies, observational bandwidth and integration time.
Even more importantly, the three 
synthetic maps demonstrate that the 21~cm line tomography can
provide a superb tool to track the redshift evolution of the reionization process (the accuracy of this
measurement could be noticeably enhanced by cross-correlating the 21~cm emission with CMB secondary anisotropies, 
see Salvaterra et al. 2005).

We have repeated the same analysis halving (doubling) the observational bandwidth, i.e. 
$\Delta \nu_{obs} =64\,\,(256)$~kHz; this corresponds to a sensitivity decrease (increase)
by a factor $\sqrt{2}$ with respect the case analyzed above. 
The brightest structures in the synthetic maps at $z=10.6$ have $\delta T_b = 34$~mK for 
$\Delta \nu_{obs} =64$~kHz with a dynamical range of 28~mK. The worse sensitivity (14 mK)
obtained for this bandwith results only in a 2-$\sigma$ detection.   
For the largest $\Delta \nu_{obs}=256$~kHz, instead the peak $\delta T_b = 33$~mK, and the dynamical range is 
21~mK. Hence, given a sensitivity of $\approx 7$~mK, this would represent a 3-$\sigma$ detection in analogy with the $\Delta \nu_{obs}=128$~kHz case. Thus, we conclude that using the median observational bandwidth is preferable 
as it allows to have a fine-grained IGM tomography without missing too much signal. 

The situation is less promising in early reionization models as the L20 we have studied. In this case, 
the rapid disappearance of the neutral IGM component negatively affects the detection in two ways: 
(i) the instrumental sensitivity decreases rapidly at low frequencies (i.e. towards higher redshift);
(ii) the sky brightness temperature increases drastically.
In this case, for a $\Delta \nu_{obs}=128$~kHz (corresponding to a sensitivity
of about 80~mK), we find a peak value $\delta T_b=56$~mK, a dynamic range of 26~mK. This
corresponds to 0.32 $\sigma$, implying an extremely difficult detection of the neutral IGM in 
an early reionization scenario.    

Thus we conclude that, although one cannot exclude that even early reionization scenarios may lead to a positive
LOFAR detection, the most favorable case would be one in which the EoR took place relatively recently.
If this is the case we have shown that in principle LOFAR will be able to see reionization in its full action. 

\begin{figure}
\begin{center}
\includegraphics[width=8cm]{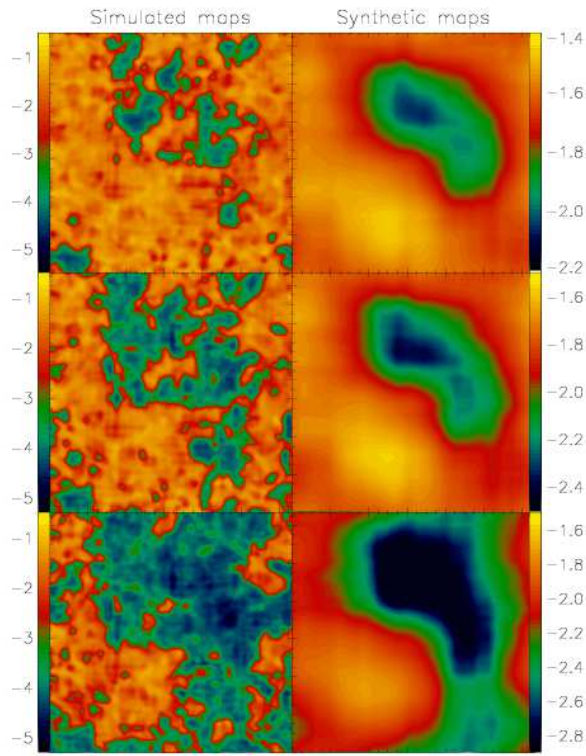}
\end{center}
\caption{Logarithmic brightness temperature, $\log[\delta T_b/K]$, maps (linear size $\approx 11$~arcmin) of the 21~cm emission 
for the late reionization model S5 at redshifts $z=10.6, 9.89, 9.26$ from top to bottom, 
respectively. Left panels: maps obtained directly from the simulation, i.e. before convolution 
with LOFAR characteristics; Right: LOFAR synthetic maps.}  
\label{graph3}
\end{figure}


\section{Conclusions}

By convolving simulations of cosmic reionization and 21~cm line emission
with the provisional characteristics of the radio telescope LOFAR, we
have shown that LOFAR has in principle the sensitivity to detect the signal from
neutral hydrogen in the IGM, at least for a late reionization scenario
(EoR at $z \approx 8$) that appears to be more appropriate after the release of the 
third year WMAP data analysis results. If
reionization occurred earlier, the experiment is much more challenging. Nevertheless,
even an upper limit on the neutral fraction at different epochs would represent a
cornerstone result which could also strongly constrain more complex/double reionization 
histories. 
Our study is an 
attempt at making specific predictions for LOFAR based
on state-of-the-art simulations, and it is affected by a number of limitations and approximations. 
Although our simulation box is one of the largest used so far in numerical reionization studies, 
the considered volume of (30~Mpc)$^3$ is only marginally able to capture the
global characteristics of cosmic reionization.
Larger boxes could be used, but information on
small scale structure (clumping) is easily smoothed out and lost, resulting 
in an extremely coarse description of the recombination processes. It is only very recently (Kohler, Gnedin 
\& Hamilton 2005, Iliev et al. 2006) that dedicated investigations has started to tackle this very difficult problem.

It is beyond the scope of this paper also to deal with the crucial
problems of ionospheric scintillation and foreground contamination
(i.e. Galactic free-free and synchrotron emission, unresolved
extra-galactic radio sources, free-free emission from ionizing sources,
synchrotron emission from cluster radio halos and relics); rather, we
consider the idealized condition in which the reionization signal is
the only one present in the sky.
A number of studies have discussed these complications in some
detail (e.g. Shaver et al. 1999; Oh \& Mack 2003; Di Matteo,
Ciardi \& Miniati 2004) and we refer the reader to those papers
for more information.
We note however, the success of the experiment will depend on the capability
of effectively removing the foregrounds and of correcting for
ionospheric variations (e.g. Gnedin \& Shaver 2004; Zaldarriaga,
Furlanetto
\& Hernquist 2004; Santos, Cooray \& Knox 2005; Morales, Bowman \& Hewitt
2005).

\section*{Acknowledgments}

We would like to thank the referee N. Gnedin for insightful comments, L. Testi and N.M. Ramanujan 
for useful discussions and J. Noordam for
providing the provisory LOFAR station configuration.

\end{document}